\documentclass[12pt]{extarticle}
 \pdfoutput=1
\usepackage{amsmath}
\usepackage{amssymb}
\usepackage{graphicx}
\usepackage[lmargin=1 in,rmargin=1 in,tmargin=.5 in,bmargin=.7  in]{geometry}

\usepackage{bm}

\newcommand{\qdot}{\dot{ q_1}}
\newcommand{\rdot}{\dot{ q_2}}
\newcommand{\sn}{\mathrm{sn}}

\newcommand{\ml}{\mathfrak{l}}
\newcommand{\Gr}{\mathcal{G}}
\newcommand{\F}{\mathcal{F}}
\newcommand{\sq}{\simeq}
\newcommand{\bt}{ \bm\tilde}

\newcommand {\la}{\langle}
\newcommand{\vp}{v_\perp}
\newcommand{\ra}{\rangle}
\newcommand{\Dv}{\Delta v}
\newcommand{\Dp}{\Delta p}
\newcommand{\lb}{\{}
\newcommand{\rb}{\}}

\begin{document}

\begin{center}
{\large
 Geometry of Thin Films 
} \\ 
${\ }$\\
Ron Perline, Department of Mathematics  \\
Drexel University, Philadelphia, PA 19104 \\
email: ronald.k.perline@drexel.edu \\
(Submitted to the Journal of Nonlinear Science, 9.17.2015)
\end{center}

\title{
Thin Films
}

\begin{abstract}
We study ray optics in the context of double mirror systems, in the limit as the two mirrors approach
 one another (thin films).  This
leads to a  novel set of differential equations on a mirror surface  which have interesting structure as seen from
 the perspective of symplectic geometry and Hamiltonian mechanics.
\end{abstract}

\begin{section}{Introduction}
In optics, the phrase "thin films" refers to optical materials for which one dimension is so much smaller than the other two that it affects the physics of light transmission
in the film medium.
For technological reasons (the difficulty of actually producing such films), the subject was not investigated until the early 1920's (see \cite{Hea} for a 
historical background prior to the 1960's).  By now, there is an enormous literature devoted to the physical properties of 
such films, optical and otherwise (for a recent summary, see \cite{Mac}).   In the design of such thin films, one can consider inhomogeneities in the film material itself; or 
the film can be constructed from a homogeneous material, but with varying thickness. Many  experimental studies of the latter case have been conducted,
going back as far as the early 70's (\cite{Ri}, \cite{Ul}).

Depending on the thickness of the film, light transmission can be modelled using the ray optics approximation, with the light rays "bouncing" from top to bottom;  these two surfaces acting like nearby reflecting surfaces (mirrors).  

Our interest in optics of thin films was stimulated by a question posed by S. Tabachnikov regarding the relationship between symplectic geometry and mirrors.   We present his problem using the language of symplectic geometry and hamiltonian mechanics, and show the connection to ray optics in thin films.
 These considerations will lead to a novel set of differential equations associated with a surface embedded in $R^3$, or more generally, hypersurfaces.  For appropriate choice of parameters, one can obtain billiard dynamics and geodesic flow as special cases of these equations.  

Consider $V = R^{n} \oplus R^{n}$, endowed with the standard inner product.
Via this inner product, we can identify $T^*S^{N}$, the cotangent bundle of the sphere, with its tangent bundle, and with the set 
$$G_1 = \{ (x,y) \in V \mid \langle x,x\rangle  = 1 , \langle x,y\rangle  = 0 \}$$

We can use $G_1$ to parameterize $\Gr$, the Grassmannian
of oriented lines in $R^n$, in the following manner:  given  an oriented line $\ml \in \Gr$,
we associate to it its unit direction vector $x = x(\ml)$, and the unique vector along
$\ml$, $y = y(\ml)$, such that   $\langle x,y\rangle =0$.

As a result of this identification, $\Gr$ inherits a symplectic structure from $T^*S^{n-1}$. We mention a few references which can be useful here: for general information about symplectic geometry, see \cite{Ab}; for specific 
information about the geometry of $\Gr$, see \cite{Marle}; a discussion of the symplectic geometric approach
to a related topic,  {\it thin lens} theory, can be found in \cite{Guil}.

Given a surface $M$, it induces (via reflection) a mapping on $\Gr$, and it is  
known that
this is
a symplectic map.  Clearly, not all symplectic maps can realized in this manner, which leads to a question of Tabachnikov's: which ones are?  One can obviously expand the question to include multi-mirror systems, in particular pairs of mirrors, and this is the particular case we will address: what are the symplectic maps on $\Gr$ induced by double mirror systems.
 
At the moment, we cannot answer this question.  However, we can make progress
on a  simpler, `` infinitesimal "
version of it.  To explain this, we need a few definitions.

Given an orientable surface $M_0$, let $\Phi_{M_0}$ be the associated symplectic map on $\Gr$.  We consider 
surfaces nearby $M_0$ as graphs in the normal bundle of $M_0$; let $f: M_0 \rightarrow R$ (we will refer to $f$ as the {\it displacement function}) and
use this to construct a one-parameter family of surfaces
 $M_\epsilon = \{ p + \epsilon f(p)N(p), p \in M_0\}$, where $N(p)$ denotes the unit normal.
With $\Phi \equiv \Phi_{M_0}$, and $\bt \Phi \equiv \Phi_{M_\epsilon}$, we consider the composition
${\Phi}^{-1} {\bt \Phi}$, corresponding to the double mirror system consisting of $M_0$
and $M_\epsilon$.  Note that, for any mirror $M$, $\Phi_M$ is an involution.

   For $\epsilon$ small, this composition should be close to the identity map,
and in fact one expects (since ${\Phi}^{-1} {\bt \Phi}$ is symplectic) that 
$\chi_f={{d} \over {d \epsilon}}( {\Phi}^{-1} {\bt \Phi} )\mid_{\epsilon = 0}$                                     
 should be a hamiltonian vector field on $\Gr$, with some Hamiltonian $H_f$. 
 So the infinitesimal version of Tabachnikov's question for double mirror systems is:  what distinguishes the
Hamiltonians $H_f$ among general Hamiltonians on $\Gr$? Note: we write "hamiltonian" if the word
is being used as an adjective, and "Hamiltonian" if it is being used as a noun.  For obvious reasons, we will
refer to these systems which are limiting versions of double mirror systems as {\it thin films}.

To most easily express a formula for $H_f$, we need a different parameterization of $\Gr$ which
is better suited to our context. Suppose we have a surface $M$; given an oriented line
 $\ml \in \Gr$, we consider (as before) the unit  direction vector $v =v(\ml)$, as well as
"the" intersection point $p = p(\ml)$ of $\ml$ with the surface $M$ (the existence and uniqueness
of $p$ depends of course on the geometry of $M$ and relative positioning of $v$).  In other
words, we are parameterizing $\Gr$ by the set $G_2 = S^{n-1} \times M \subset V$.  We
have the map $T: G_2 \rightarrow G_1$ defined by $(v,p) \rightarrow (v, p - \langle p,v\rangle  v)$, which
allows one to pull back the symplectic structure from $G_1$ to $G_2$.   We will freely 
identify points in $G_2$ with corresponding points in $\Gr$.

The vector field $\chi_f$ can, via the above identifications, be considered as a vector field on
$G_2$.  {\it Then $\chi_f$ corresponds to the Hamiltonian $$H_f(v,p) = -2\langle v,N(p)\rangle f(p) \enskip . $$} 

The formula for $\chi_f$ is given by Eqn. (1) of the next section.  

Naturally, this raises the following question:  for a given $H$ on $\Gr$, how do we know there
exists an $M$ and $f$ so that $H =H_f$? This question is not yet solved; however,
it is likely 
that there is some differential condition which needs to be satisfied.  Notice, for example, in the form 
given above, that the differential of $H_f$ with respect to the $v$ variable is independent of $v$.

The structure of this paper is as follows:  In Section 2, we explicitly obtain the formula for $\chi_f$ on $G_2$.
In Section 3, we show that the Hamiltonian indicated above does indeed correspond to $\chi_f$.
In Section 4, we discuss the dynamical interpretation of $\chi_f$ and its relation to multiple reflections.
Sections 5 and 6 show connections between this dynamical problem, geodesic flow, and billard dynamics.
Section 7 gives another formulation of our equations coming from a related variational problem, which can be more convenient to work with.
 We then give a fairly complete discussion to one concrete problem: the dynamical problem associated with a spherical displacement function. 
This is followed by a brief discussion of the dynamical problem associated with an ellipsoidal displacement function.
Section 8 considers ``error estimates" comparing the dynamical problem associated with $\chi_f$ and the discrete system of multiple reflections inside the thin film.
Section 9 describes some bracket relations for the types of
Hamiltonians we are considering: it indicates that there is a natural Lie algebra (generated by the Hamiltonians $H_f$) which includes a subalgebra
isomorphic to the vector fields on the mirror surface (with the standard Lie bracket).  Section 10 contains some concluding remarks.

The goal of this paper is to introduce $\chi_f$ and its various basic properties.  We briefly make some comments about some cases where the associated dynamical problem is integrable; a fuller treatment of more extensive integrable cases will be treated in a subsequent publication.

\end{section}

\begin{section}{Computation of $\chi_f$ on $G_2$}
The procedure for the explicit computation of $\chi_f$  is graphically indicated in Figure 1. We give a brief summary of the steps involved.  First a remark:  we will need to compute the objects and quantities 
mentioned below modulo terms of order $\epsilon^2$, and  we will write  $A \sq B$ if  $A-B = O(\epsilon^2)$. In the following,
``compute" will mean "compute modulo terms of order 2 or higher in $\epsilon$.  Hopefully, this will not cause confusion.

Let's fix $\epsilon$ small, and denote $M = M_0, \bt M = M_\epsilon$; $\Phi , \bt \Phi $ are as before.
We have discussed the map $\Phi$, but have not yet described it explicitly.  Let $\ml$ be a line in $\Gr$
represented by a point $(v,p) \in G_2$, then 
$\Phi(v,p) = (v_r, p)$, where $v_r = v - 2\langle v, N(p)\rangle N(p)$.
"Above" each point $p \in M$ is a corresponding point $\bt p = p + \epsilon f(p) N(p)$ ($f$ could of course be negative).
 At $\bt p$ there is $\bt N ( \bt p)$  which is normal to $\bt M$. 
We will need to compute $\bt N$ in terms of $M$ and $f$.  

To compute $\bt \Phi$ applied to a line $\ml$, we will need the point of intersection of $\ml$
with $\bt M$, which we call $\bt i$, as well as the normal $\bt N(\bt i)$ at that point.
$\bt i = i +\epsilon f(i)N(i)$ for some point $i$ on $M$; knowing $i$ will allow us to compute
$\bt N ( \bt i)$.  With this information, we can compute $\bt v_r = v - 2 \langle v, \bt N (\bt i)\rangle  {\bt N(\bt i)}$.

To complete the evaluation of $\bt \Phi(\ml)$, we need to find the point of intersection $q$ of the reflected line with the
surface $M$.

Finally, our goal is to compute $\Phi^{-1} \bt \Phi(\ml)$.  To do so, we will need to reflect $\bt v_r$  at the point $q$;
$w =\bt v_r - 2\langle  \bt v_r, N(q)\rangle N(q)$.  Thus, $\Phi^{-1} \bt \Phi(\ml)$ has coordinate representation 
$(w, q)=(w(\epsilon), q(\epsilon))$ in $G_2$.  We emphasize: all the quantities and terms described above are computed modulo $\epsilon^2$.
At this point, we will compute ${{d}\over{ d \epsilon}}(w,q)|_{\epsilon = 0}$, giving us the desired vectorfield $\chi_f$ on $G_2$.

Although elementary, the calculation is somewhat tedious; the details will appear in the published version of this paper.  After these
calculations, one obtains:

 {\it The variation vector field $\chi_f(v,p)$ on $G_2$ is given by}
\begin{align}
-2 \, ( \ \la v, -uW + \nabla f \ra N - \la v, N \ra (-uW + \nabla f) \, , \,  -u \pi(v)  \ ) \, ,
\end{align}
where the dependencies $W = W(\pi(v)), f =f(p), N = N(p)$ are surpressed for readability, and $u = { {f(p)} \over {\la v, N(p) \ra }}$ as before; $W$ denotes the Weingarten map for the surface $M$.


\end{section}



\begin{section}{The Symplectic Structure and the Hamiltonian}
In addition to an inner product,  $V = R^{n} \oplus R^{}$ has a symplectic structure:  if $w_1=(v_1, p_1), w_2=(v_2, p_2) \in V$, then $\Omega_1(w_1, w_2) =
\la v_1, p_2 \ra - \la v_2, p_1 \ra$.
As a result of a simple calculation (or obviousness) the symplectic structure on $G_1$ that one obtains via the identification of $G_1$ with $T^*S^{n-1}$ is just the restriction 
of $\Omega_1$ to $G_1$.  Let's now discuss the symplectic structure on $G_2$.  In the introduction, we defined a map $T: G_2 \rightarrow G_1$, which can be extended to
a mapping from $V$ to $V$.  We can use this to pull back
the symplectic structure on $G_1$ to $G_2$.  This is the same considering $T$ as a map from $V$ to $V$,  computing the pullback of $\Omega_1$ via $T$ (call this $\Omega_2$)
and then restricting $\Omega_2$ to $G_2$.  $\Omega_1 \neq \Omega_2$ on $V$, but the two forms {\it  are} equal when restricted to $G_2$.  Let's show this.
From the definition of $T$, considered as a map on $V$ one computes that 
$$dT(\Delta v, \Delta p)  =
 (\, \Delta v \, , \, \Delta p - \la \Delta p, v \ra v - \la p, \Delta v \ra v - \la p , v \ra \Delta v  \, ) \ .$$
This can be conveniently written in matrix form:
$$
 \begin{bmatrix} I_n & 0_n\\ {{(p^\dagger v)} \,  I_n - v p^\dagger} & {I_n - v v^\dagger} \end{bmatrix}  \left[ \begin{array}{c} {\Delta v}  \\ {\Delta p} \end{array} \right] \ .
$$
Abusing notation, we will refer to this $2n \times 2n$ matrix as $dT$.   $\Omega_1$ can be represented by the matrix $J$, with
$$J = \begin{bmatrix} 0_n & I_n\\ -I_n  & 0_n\end{bmatrix} , $$
and so the matrix representation of $\Omega_2 = dT^\dagger J dT$ is
$$J_2 = \begin{bmatrix} {pv^\dagger -v p^\dagger} &{ I_n-v v^\dagger} \\{ -I_n+ v v^\dagger}  & 0_n\end{bmatrix} . $$
Of course, by this we mean the following: if $w_i = (\Delta v_i, \Delta p_i), i = 1, 2$  then $ \Omega_2(w_1,w_2) =w_1^\dagger J_2 w_2$.
Let us now restrict our attention to vectors tangent to  $G_2$, which means we are considering $\Delta v_i$ and $\Delta p_i$ with $\la \Delta v_i , v \ra = 0$ and
$\la \Delta p_i , N(p) \ra = 0$. Under these conditions, we have  $w_1^\dagger J_2 w_2 = w_1^\dagger J w_2$.

Using $J$, we can compute the co-vector associated to the vector $\bt \chi_f = \frac{- \chi_f}{2}$;  application of this co-vector to a vector  is just the result of the 
matrix multiplication
$$
\begin{bmatrix} (\Delta v)^\dagger  & (\Delta p)^\dagger  \end{bmatrix}  \begin{bmatrix} 
  -u \pi(v) \\ - \la v, -uW + \nabla f \ra N + \la v, N \ra (-uW + \nabla f) \end{bmatrix} \, .$$
Let's look at the term involving $\Delta v$.  We obtain:
$$\begin{array} {lcl}
A = \la \Delta v, -u \pi (v) \ra & = & -u \la \, \Delta v, v -\la v, N \ra N \,  \ra\\
& = & -u \la \Delta v, -\la v, N \ra N \ra \\
&=&  u \la v, N \ra \la \Delta v, N \ra \\
&=& \la \Delta v, N \ra f(p) \, . \\
\end{array}
$$
Treating the terms which include $\Delta p$, we have
$$
\begin{array} {lcl}
B&= &      \la  \, \Delta p ,    - \la v, -uW + \nabla f \ra N + \la v, N \ra (-uW + \nabla f) \, \ra \\
&=&   \la  \, \Delta p ,    \la v, N \ra (-uW + \nabla f) \, \ra \\
&=& -u \la v , N \ra \la \Delta p , W \ra + \la \Delta p, \nabla f\ra \la v, N \ra \\
&=&   \la \Delta p, -W \ra f(p)  +  \la \Delta p, \nabla f\ra \la v, N \ra  \,  . \\
\end{array}
$$

Now, consider the Hamiltonian $ \bt H_f(v,p) = \la v, N(p) \ra f(p)$.  We calculate its differential:
$$
\begin{array}{lcl}
d  \bt H_f (\Delta v, \Delta p) &=& \la \Dv, N \ra f(p) + \la v, - W(\Dp) \ra f(p) + \la v, N \ra \la \nabla f, \Dp \ra \\
&=& \la \Dv, N \ra f(p) + \la \pi(v), - W(\Dp) \ra f(p) + \la v, N \ra \la \nabla f, \Dp \ra \\
&=& \la \Dv, N \ra f(p) + \la-W( \pi(v)), \Dp \ra f(p) + \la v, N \ra \la \nabla f, \Dp \ra \\
&=& A + B \ ,
\end{array}
$$
which allows us to conclude that $\chi_f$ is indeed the hamiltonian vector field associated with $H_f$.
\end{section}

\begin{section}{Dynamical Interpretation of $\chi_f$}
The calculations of the previous sections are related to the evaluation of ${ d \over {d \epsilon}} \Phi^{-1} \bt \Phi |_{\epsilon =0}$ Our calculation of $\chi_f$
essentially says the following:
Start at a point $(v_0, p_0)$, reflect off $\bt M$, intersect $M$ at a point $p_1$, then reflect again at $p_1$.  Call the new point in $G_2$
$(v_1, p_1)$.  Then $(v_1, p_1) \sq (v_0, p_0) + \epsilon \chi_f (p_0)$ 

One can obviously iterate this process, to obtain subsequent points $(v_i, p_i), i =  0, 1, 2, 3 \dots$. Thus, given a fixed time $T_0$, it seems that the time $T_0$ map of the flow 
associated to the equation ${{dg} \over {dt}}= \chi_f(g)$  on $G_2$ is the formal limit of the "double reflections" shown in figure 2, iterated ${T_0} \over \epsilon$ times, with spacing between surfaces of order $\epsilon$.
To prove this, one can reason along the lines of numerical analysis, where there are "local truncation errors" and "global errors" associated with numerical schemes for solving ODE's (see section 10).  
\newline
Given this dynamical interpretation, we will refer to the differential equation defined on $G_2$ given by $\dot \gamma(v,p,t) = \chi_f$ 
(here, "$\dot \gamma $" denotes differentiation with respect to $t$) as the ``Thin Film Equation" (TFE). In the next two sections, we consider two simplified cases:
 one,  where $M$ is an arbritrary (in particular, curved) surface in $R^{N+1}$ and the displacement function is constant;
 and the second  is  where the surface $M$ is a plane, and $f$ is an arbitrary displacement function.

 For the most part, we will be concentrating on $N = 2$ ($M$ is a surface in 3-space)  A note on terminology:  the vector $v$, since it has a normal component, is in general not in the same direction as the tangent vector to the $p$ trajectory; so let's call it the pseudo-velocity.
\end{section}

\begin{section}{Relation to Geodesic Flow}

The situation where $M$ is not planar, and  the film is uniform ($f \equiv c$)  is interesting; here $H(v,p) =c \la v, N(p) \ra$.  As we shall now show, the associated film equations are essentially a scaled version of geodesic flow. 

 In the case $f = c$,, the $\nabla f$ terms in $\chi_f$ disappear, and we are left with
$$u \, ( \ \la v, W(\pi(v))  \, \ra N(p) - \la v, N(p) \ra W(\pi(v))  \, , \,   \pi(v)  \ ) \, $$
Note  that ignoring the factor $u$ (which is essentially the reciprocal of the Hamiltonian!) simply causes a change of time scale for the
 solution to the differential equation, and we will therefore
discard it.
We have thus arrived at the differential equation system on $G_2$,
\begin{align}
{{dv} \over {dt}} = \la v, W \ra N - \la v, N \ra W  ,  \ \ { {dp} \over {dt}} = \pi \ \,  ,
\end{align}
where for the moment we use the following abbreviated notation:  $W \equiv W(\pi(v))$, $N \equiv N(p)$, $ \pi \equiv \pi(v)$, $\vp \equiv \la v, N \ra N$. Note that 
$\alpha \equiv \la v, N \ra$ can be treated as a constant.  Thus we have:
$$ v = \pi + \vp, $$
$${{dv} \over {dt}} = \la \pi, W \ra N - \alpha W,$$
$${{dv} \over {dt}} = {{d\pi} \over {dt}} +  {{d\vp} \over {dt}}.$$
Now since $\vp = \alpha N$, we have
$$ {{d\vp} \over {dt}} = \alpha  {{d N} \over {dt}}= -\alpha W .$$
Using this last fact, and the two equations given above involving 
$ {{dv} \over {dt}}$, we obtain

$${{d\pi} \over {dt}} = \la \pi, W \ra N , \, \,  { {dp} \over {dt}} = \pi$$
One sees easily that $\la \pi, \pi \ra$, is constant, so we can rescale $t$  to arclength
$s$ so that ${{dp} \over {ds}} = T$, where $T$ is the tangent indicatrix. The equations
then become 
$${{d T} \over {ds}} = \la T, W(T) \ra N , \, \,  { {dp} \over {ds}} = T$$
which are the equations for a geodesic on a (hyper)-surface.
 Thus, our equations (with general $f$) generalize the notion of
light rays on a surface.
\newline
\end{section}
\begin{section} {Flat Substrate Films and  Billiards} 
Consider the situation where $M$ is the plane (so $W=0$), and $f= f(p) = f(p_1,p_2) $ is some arbitrary function. Then, from the last section, we can visualize the dynamics of the TFE as a ``smoothed out" version of the discrete system obtained by a particle (or light ray) bouncing up and down between a flat plate substrate and a nearby, nearly flat contour. Then of course,  the differential equation associated with $f$ looks like (we ignore the factor of $-2$ and rescale by a factor of $\la v , N \ra$, which is {\it not} the Hamiltonian here)
\begin{align}
 \frac{dv}{dt} = \la v, N \ra \la v, \nabla f \ra N - \la v , N \ra^2 \nabla f, \, \,   \frac{dp}{dt} = - f(p)\pi(v)   \, .
\end{align}

To get a feel for some of the possible dynamics, consider the plots in Figures 3-6.  The caption for each plot gives the displacement function $f$, and the initial values of 
$v$ and $p$ respectively (for these plots, $p_1$ and $p_2$ refer to coordinates in the plane).  Of course, we are simply looking at the $p$ evolution; the $v$ evolution is taking place on the unit sphere.  In some of the examples, there seems to be interesting regularity to the solution path; but at least one of the pictures seems to show rather chaotic behavior.

Consider Figures 5 and 6.  The displacement function and initial position are the same; only the inclination (initial $v$) is different.  for Figure 6, the $v$ is nearly horizontal.  The two trajectories (or more precisely, their projection onto the $p$ coordinates) in both cases lie within the unit circle, but the second is somewhat striking it that the trajectories look like billiard paths (straight) and, at least from the graphic, it looks like the reflection angle is the appropriate one for Snell's law to hold.  Here is a simple heuristic argument as to why this should be the case.

First of all, it is worth looking at the simple displacement function 
$f(p_1, p_2) = -p_1$.  In this case, the relevant differential equations are:
$$
\begin{array}{lcl}
\frac{d p_1}{dt} &=& p_1 v_1 \, , \\
\frac{d p_2}{dt} &=& p_1 v_2  \, ,  \\
\frac{d v_1}{dt} &=& {v_3}^2 \, , \\
\frac{d v_2}{dt} &=&  0 \,  , \\
\frac{d v_3}{dt} &=& - v_3 v_1
\end{array}
$$
The equations are simple enough so that they can be solved explicitly: assuming the that third coordinate of the the initial vector $v$ is positive, and that the initial position $p$ has negative horizontal component, then one can show that the resulting trajectories are of the form
$$ \left( {\it p_2}-A \right) ^{2}-{B}^{2}{{\it p_1}}^{2}=-{C}^{2} \, ,$$
a hyperbola whose axis of symmetry is horizontal (and hence perpendicular to the line
$f = -p_1 = 0$.  (See  Figure 7).

This easily extends to the general linear case: If $f(p_1, p_2) = a p_1 + b p_2$;
and if we assume a starting point $p$ with $f(p) \ge 0$ and initial vertical velocity component positive, 
then the associated solutions curves are hyperbolae whose axis of symmetry is parallel to 
$\nabla f = [a, b]$.

Now, consider an $f$ with the following properties:  $f$ is regular at the points $p$ where
$f = 0$; that the regular curve $\Gamma$ defined by $f(p) = 0$ is convex and bounded; and that the function is positive on the interior of the regular curve.  Let's start off with a nearly horizontal initial pseudo-velocity vector $v_0 = (v_1, v_2, \epsilon)$, where $\epsilon$ is a small positive constant.   If we consider the system Eqn (2), we see that $v$ is changing extremely slowly (and in particular, the rate of change of the tangential part of $v$ is of order $\epsilon^2$), so for a period of time the direction of the velocity of $p$  is nearly constant even if the speed is varying; hence the trajectory of $p$ is nearly a straight line.  However, as we approach a point $\gamma$ of the boundary curve $\Gamma$, and $f(p)$ approaches zero, the dynamics mimics the behavior of the trajectories associated with the function $g$ which is the linearization of $f$ at $\gamma$; the trajectories of $g$ look nearly like straight lines which reflect as in Snell's law.  Our informal discussion leads to the conclusion that {\it as the $v$ vector becomes horizontal, trajectories of Eqn (3) approach the orbits of the billiard problem associated with the curve $f$ = 0}.  Of course, there is another method for obtaining billiard dynamics from a related smooth dynamical system:  the  Birkhoff construction
 of looking at geodesic flow on ``pancake"-like surfaces, and letting the thickness of the surfaces  go to zero (see \cite{Ta})  We observe that the two constructions are quite different
\end{section}
\begin{section}{Variational formulation}
In this section,  we work with flat templates.  We describe how the thin film equations are related to a variational problem,
 which for certain examples simplifies the solution of the thin film equations.
We then give two examples of thin films on flat templates, with displacement functions 
such that the associated thin film equations are integrable.  We will discuss the first (and easier one) in some detail; we examine the second one only briefly.

\begin{subsection}{Variational formulation}
Define $N = n-1$.  The thin film equations for a flat template in $R^{N+1}$ can be written:
$$ 
\dot{v}_{N+1} =  \sum _{k=1}^{N}v_{{k}}   {\frac {\partial f}{\partial q_{{k}}}  }; \qquad
\dot{v}_{k} = - v_{N+1}         {\frac {\partial f}{\partial q_{{k}}}  }       ,   \quad
\dot{q}_k = \frac{-f}{v_{N+1}}v_k,  \quad k = 1 \dots N
$$
Note that $\sum_{ }^{N+1} \dot{v}_j v_j = 0$, so the condition $\sum^{N+1}{v_j}^2 = c$ is preserved by the flow.  Without loss of generality, we can assume $c=1$.

If we define $D \equiv \sqrt{ 1 - \sum^N {v_k}^2 }$, then we can rewrite our equations with $v_{N+1}$ eliminated, obtaining the reduced form:
$$ \dot{v}_k = -{D}   \,\,  {\frac {\partial f}{\partial q_{{k}}}  } , \quad      \dot{q}_k = -\frac{f}{{D}} v_k, \qquad k = 1 \dots N$$
Now consider curves which lie in the hyperplane $v_{N+1} = 0$, with functional:
$$
\F  = \int F(q_1, \dots q_n, \dot{q}_1 \dots , {q}_N )\,\, dt ,    \qquad    F =  \sqrt{ \sum^N (\dot{q}_k)^2 + f^2(q_1, q_2, \dots q_n) } \,\,
$$
Of course, $F$ is the Lagrangian.  The procedure for computing the associated Hamiltonian and Hamilton's equations is routine (\cite{Gel}) Define
$$ p_k = \frac{\partial F}{\partial \dot{q}_k} = \frac{1}{{D}} \dot{q}_k \, , $$
which gives $p_k$ in terms of $\dot{q}_k$.  We need to solve the equations for $\dot{q}_k$ in terms of $p_k$; a bit of algebra leads to
$$
\dot{q}_k  = \frac{p_k f}{E}, \quad E = \sqrt{ 1 - \sum^N p_j^2}; 
$$
one can check directly that these formulas work.

The Hamiltonian is $H = F - \sum^N \dot{q}_k p_k$, which translates into (now that we have $\dot{q}_k$ in terms of $p_k$) $H = f E$.  Then one computes Hamilton's equations
$$
\dot{q}_k = \frac{ \partial H}{\partial p_k}, \qquad \dot{p}_k = - \frac{\partial H}{\partial q_k},
$$
which for our problem become
$$\dot{q}_k = -\frac{ f p_k}{E}, \quad \dot{p}_k = - \frac{\partial f}{\partial q_k} E$$
\end{subsection}
which are exactly the reduced thin film equations given above, with $v_k$ replaced with $p_k$. Making this identification, we 
observe that $H = f E = f \sqrt{ 1 - \sum^N p_j^2} = f \sqrt{ 1 - \sum^N v_j^2} = f v_{N+1} =  \langle v, N \rangle \, f; $, thus we could have ``guessed" the form of $H$.
\begin{subsection}{The Spherical Displacement Function} 
Consider a flat template in the plane, with  ``spherical" displacement function
$f(q_1, q_1) = \sqrt{1-{q_1}^2- {q_2}^2}$.   By "spherical", we simply mean that when the ``small" parameter $\epsilon$ is set equal to 1, the graph of $f$ gives a (hemi-) sphere. Then, as discussed at the beginning of this section, the associated variational problem is to find critical points for the functional 
$$ 
\F = \int {\sqrt{     { \dot{q_1}}^2 + { \dot{q_2}}^2 +  1-{q_1}^2- {q_2}^2 } \, dt }
$$
Given the symmetry of the problem, it is hardly surprising that we can solve the associated equations exactly; but on the other hand
a complete analysis of the problem can be useful in understanding more complicated problems.
It is useful to transform this into a functional on curves parameterized in polar coordinates
$q_1 = r cos(\theta)$, $q_2 = r sin(\theta)$:
$$
\F_{polar}   =  \int {\sqrt{  \,  {\dot{\theta}}^2 \,  { r}^2 + { \dot{r}}^2 +  1-{r}^2} \,  \, dt }
$$
The associated Euler-Lagrange equations are:
$$
\ddot{r} ={\frac {r \left( {{ \dot{\theta}}}^{2}{r}^{2}-{r}^{2}+2\,{{ \dot{r}}}
^{2}-{{ \dot{\theta}}}^{2}+1 \right) }{{r}^{2}-1}}, \ \  \\
\ddot{\theta} 
=2\,{\frac {{ \dot{r} }\,{ \dot{\theta}}}{r \left( {r}^{2}-1 \right) }}
$$
A plot of an example solution curve, given by numerically solving these differential equations, is given in Figure 8. The initial conditions for this example are: 
$$
\left\{ r \left( 0 \right) =(1/2),\theta \left( 0 \right) =0,
  \dot{r}  \left( 0 \right) =0,  \dot{\theta}
  \left( 0 \right) =(3/7)\,\sqrt {21} \right\} 
$$

We now continue to find explicit solutions.

Observe that the second order equation for $\theta$, is linear and easily solved in terms of $r$:
$$
\theta \left( t \right) =D + C\int \!{\frac { \left( r \left( t \right) 
 \right) ^{2}-1}{ \left( r \left( t \right)  \right) ^{2}}}{dt}
$$
For the moment, we assume $D=0$.
We can substitute this formula for $\theta$ into the differential equation for $r$, to obtain
$$
\ddot{r}={\frac {{C}^{2}{r}^{6}-3\,{C}^{2}{r}^{4}-{r}^{6}+2\,{{\dot{r}}}^{2}
{r}^{4}+3\,{C}^{2}{r}^{2}+{r}^{4}-{C}^{2}}{{r}^{3} \left( {r}^{2}-1
 \right) }}
$$
Of course, this is second order in $r$ with an auxiliary parameter $C$.  To solve this, we consider the ansatz
$$
r=\sqrt {a+b \left( {\sn} \left( ct,k \right)  \right) ^{2}}
$$
and then see what the required relations are between the constants $a,b,c,k$ required so that the ansatz does indeed give a solution.
After some manipulation, one obtains a two parameter family of solutions of the form of the ansatz:
$$
r =\sqrt { \left( \sin \left( w \right)  \right) ^{2} \left( \sin \left( 
v \right)  \right) ^{2} \left( {\sn} \left( {\frac {t}{\cos \left( 
w \right) \sin \left( v \right) }},\sin \left( w \right)  \right) 
 \right) ^{2}+ \left( \cos \left( v \right)  \right) ^{2}}
$$
corresponding to the $C$ value
$$
C=-{\frac {\sqrt {- \left( \cos \left( w \right)  \right) ^{2} \left( 
\sin \left( v \right)  \right) ^{2}+1} \, \cos \left( v \right) }{\cos
 \left( w \right)  \left( \sin \left( v \right)  \right) ^{2}}}
$$
The parameters $v,w$ range between $0$ and $\frac{\pi}{2}$.

With the explicit formula for $r$ and $C$, we can go back to finding an explicit formula for $\theta$.  The formula given above for $\theta$ involves an antidifferentiation whose integrand is an elliptic function; fortunately, this is a doable integral, and we finally obtain
$$
\begin{aligned}
\theta \left( t \right)   &=\left( \sqrt {- \left( \cos \left( w \right) 
 \right) ^{2} \left( \sin \left( v \right)  \right) ^{2}+1}\right)  \\
& \times \Pi 
 \left( {\sn} \left( {\frac {t}{\cos \left( w \right) \sin \left( v
 \right) }},\sin \left( w \right)  \right) ,-{\frac { \left( \sin
 \left( v \right)  \right) ^{2} \left( \sin \left( w \right)  \right) 
^{2}}{ \left( \cos \left( v \right)  \right) ^{2}}},\sin \left( w
 \right)  \right)  \left( \cos \left( v \right)  \right) ^{-1} \left( 
\sin \left( v \right)  \right) ^{-1}\\
 &-{\frac {\sqrt {-
 \left( \cos \left( w \right)  \right) ^{2} \left( \sin \left( v
 \right)  \right) ^{2}+1}\cos \left( v \right) t}{\cos \left( w
 \right)  \left( \sin \left( v \right)  \right) ^{2}}}
\end{aligned}
$$
where $\Pi$ (with three arguments) is defined as the incomplete elliptic integral
$$
\Pi  \left( x,c,k \right) =\int_0^x \!{\frac {dx}{ \left( -c{x}^{2}+1
 \right) \sqrt {-{x}^{2}+1}\sqrt {-{k}^{2}{x}^{2}+1}}};
$$
and $\Pi$ with two arguments is the complete elliptic integral
$$
\Pi  \left( c,k \right) =\int_0^1 \!{\frac {dx}{ \left( -c{x}^{2}+1
 \right) \sqrt {-{x}^{2}+1}\sqrt {-{k}^{2}{x}^{2}+1}}}.
$$

Since the Euler equations are second order in $r, \theta$, one expects a four parameter family of solutions for the equations.  As presented, we only have a two parameter family of solutions, which are distinguished by the fact that they satisfy the conditions
$\dot{r}(0) = 0, \theta(0)= 0$.  However, observe that our equations have two symmetries which we can exploit:  the equations are invariant under $t \rightarrow t +c,  \theta \rightarrow \theta +d$.  These two extra degrees of freedom allow us to generate all solutions to the Euler equations.

Given these explicit formula for $r$ and $\theta$, we can read off interesting geometric information about the solutions.  Consider the formula given above for $r$.  Since $\sn(x,k)$ ranges between $0$ and  $1$ as $x$ ranges between $0$ and $K(k)$ (the complete elliptic integral of the first kind), we see that $r$ ranges between $r_{min} =\cos(v)$ and 
$r_{max}=\sqrt { \left( \sin \left( w \right)  \right) ^{2} \left( \sin \left( 
v \right)  \right) ^{2}+ \left( \cos \left( v \right)  \right) ^{2}}$.  
In fact, $r(0) = r_{min}$ and $\dot{r}(0) = 0$.  Also, $\theta(0) = 0$ and 
$\dot{\theta}(0) ={\frac {\sqrt {- \left( \cos \left( w \right)  \right) ^{2} \left( 
\sin \left( v \right)  \right) ^{2}+1}}{\cos \left( v \right) \cos
 \left( w \right) }}
 $.  This allows us, by appropriate choice of $v,w$ to construct solutions with $r_{min}, r_{max}$ as desired.  For example, for the choice 
$r_{min} =1/2, \, r_{max} = 3/4$, one readily checks that $v= \frac{\pi}{3}, \, w=\arcsin \left( (1/6)\,\sqrt {15} \right) $;  and for this choice of the pair $v,w$, we have $\dot{\theta}(0) = (3/7)\,\sqrt {21}$, which gives us exactly the initial conditions for our numerically computed example.  We now know that our solution curve is "trapped" between the two circles of radii $1/2$ and $3/4$ respectively.  In Fiigure 9 the same solution curve, these two circles, and the circle of radius $1$ are plotted simultaneously. The initial conditions were chosen so that the inner and outer radii have simple values.

Figure 9  suggests that there is a periodic orbit nearby with five "lobes".  Indeed, resorting again to the explicit formula for $r$ and $\theta$ we can find closed solutions (periodic orbits) for our equations.

First, let's consider the periodicity of $r$.  Since $\sn(x,k)^2$ is $2 K(k)$-periodic, then we can see that $r$ has period
$2\,{K} \left( \sin \left( w \right)  \right) \cos \left( w \right) \sin \left( v \right) 
$.  Over this interval, one checks that $\theta$ is changed by an increment of

$$
\begin{aligned}
\Delta= &-2\,\sqrt {- \left( \cos \left( w \right)  \right) ^{2} \left( 
\sin \left( v \right)  \right) ^{2}+1} \\
 &\times \left( K \left( \sin \left( w
 \right)  \right)  \left( \cos \left( v \right)  \right) ^{2}-\Pi
 \left( -{\frac { \left( \sin \left( w \right)  \right) ^{2} \left( 
\sin \left( v \right)  \right) ^{2}}{ \left( \cos \left( v \right) 
 \right) ^{2}}},\sin \left( w \right)  \right)  \right)  \left( \cos
 \left( v \right)  \right) ^{-1} \left( \sin \left( v \right) 
 \right) ^{-1}
\end{aligned}
$$

Thus, to obtain a closed curve, we need $\frac{\Delta}{2 \pi}$ to be rational.  Suppose, for example, we wish to find a five-lobe solution to our equations; we need  $\frac{\Delta}{2 \pi}= \frac{2}{5}$.  We look for a solution close to the non-periodic solution plotted in Figure 9; so we again set $v = \frac{\pi}{3}$, thus setting $r_{min} = 1/2$.  Then, we are asking for a solution to
$$
\begin{aligned}
 \frac{4 \pi}{5} =&-2/3\,\sqrt {-3/4\, \left( \cos \left( w \right)  \right) ^{2}+1} \,
(\sqrt {3}{K} \left( \sin \left( w \right)  \right) )\\
&+8/3\,
(\sqrt {-3/4\, \left( \cos \left( w \right)  \right) ^{2}+1} )\, \sqrt {3} \, {\Pi} \left( -3\, \left( \sin \left( w \right)  \right) ^{2}
,\sin \left( w \right)  \right)  
\end{aligned}
$$
and one obtains numerically $w= 0.6281734812$. Note that this not only gives closure; having this value allows us to compute $r_{max} = 0.7134529902$, compared to 
$r_{max} = .75$ for our previous example.

We now look at this example directly from the variational point of view (rather than looking at the Euler equations directly).  In rectangular coordinates,  the integrand of the functional (the Lagrangian)
$
F = {\sqrt{     { \dot{q_1}}^2 + { \dot{q_2}}^2 +  1-{q_1}^2- {q_2}^2 }  }
$.  This has two symmetries that we exploit using Noether's theorem.  The first is the independence of the Lagrangian with respect to the variable of integration $t$; the second is the invariance with respect to rotation.  The first leads to the "constant of motion"
$ H =F - \dot{q_1}F_{  \dot{q_1}}   - \dot{q_2}F_{  \dot{q_2}} $, and the second leads to the constant of motion
$ L = F_{\qdot}q_2 - F_{\rdot}q_1 $, or equivalently
$$
H = -{\frac {{q_{{1}}}^{2}+{q_{{2}}}^{2}-1}{\sqrt {{{\qdot}}^{2}+{{\rdot}}^{2}-{q_{{1}}}^{2}-{q_{{2}}}^{2}+1}}}, \qquad
L = {\frac {{\qdot}\,q_{{2}}-{\rdot}\,q_{{1}}}{\sqrt {{{\qdot}}^{
2}+{{\rdot}}^{2}-{q_{{1}}}^{2}-{q_{{2}}}^{2}+1}}}
$$
Of  course, these can both be converted into polar coordinates;
$$
H =
-{\frac {{r}^{2}-1}{\sqrt {{r}^{2}{{\dot \theta}}^{2}-{r}^{2}+{{\dot r}}^{2}+1}}}, \qquad
L =
-{\frac {{r}^{2}{\dot \theta}}{\sqrt { \, {{\dot \theta}}^{2}{r}^{2}
+{{\dot r}}^{2}-{r}^{2}+1}}}
$$
We now give an informal argument showing that $r_{min}$ and $r_{max}$  can be expressed in terms of these conserved quantities $H$ and $L$.  
Suppose we are at a point where $\dot r = 0$ (for example, when $r= r_{min}$ or $r = r_{max}$). Then the above formulas simplify:
$$
{H}=-{\frac {{r}^{2}-1}{\sqrt {{{\dot \theta}}^{2}{r}^{2}-{r}^{
2}+1}}},   \qquad {L}=-{\frac {{r}^{2}{\dot \theta}}{\sqrt {{{\dot \theta
}}^{2}{r}^{2}-{r}^{2}+1}}}
$$
So if we are at a point where $\dot r = 0$, then we can solve for the values of $r$ and $\dot \theta$ in terms of $H$ and $L$; and in particular we obtain the following formulas for $r$, which we give temporary names $U$ and $V$:
$$
\begin{aligned}
U &= 1/2\,\sqrt {2-2\,{{\it H}}^{2}+2\,{{\it L}}^{2}-2\,\sqrt {{{\it H}}
^{4}-2\,{{\it H}}^{2}{{\it L}}^{2}+{{\it L}}^{4}-2\,{{\it H}}^{2}-
2\,{{\it L}}^{2}+1}}, \\
V &= 1/2\,\sqrt {2-2\,{{\it H}}^{2}+2\,{{\it L}}^{2}+2\,\sqrt {{{\it H}}
^{4}-2\,{{\it H}}^{2}{{\it L}}^{2}+{{\it L}}^{4}-2\,{{\it H}}^{2}-
2\,{{\it L}}^{2}+1}}
\end{aligned}
$$
From the formulas, it is clear that $V > U$.  As we move along the curve, both $U$ and $V$ are constant; and at a point
where $r = r_{min}$ or $r = r_{max}$, one of these two functions gives the $r$ value at that point.  Henceforth, we
denote $U$ by $r_{min}$ and $V$ by $r_{max}$.

Obviously, the formulas above are somewhat awkward.  One can check that the inverse formulas are much cleaner;
in fact:
$$ r_{min} r_{max} = L, \qquad  { r_{min}}^2 + { r_{max}}^2 = L^2 + 1 - H^2$$
\end{subsection}
\begin{subsection}
{The Ellipsoid}
We consider the ellipsoidal displacement function 
$f = \sqrt {1-\sum _{i=1}^{N}{\frac {{x_{{i}}}^{2}}{{a_{{i}}}^{2}}}} \, $.  From Figures 11 and 12, we see a regularity of orbits for the associate TFE with $N=2$ which suggests the problem is integrable.  As a step towards substantiating this, we record constants of motion associated with this problem: define $x_{N+1} = 0, a_{N+1}= 0$, then the functions 
$$
I_k = {v_{{k}}}^{2}+\sum _{j=1, j\neq k}^{N}{\frac { \left( -x_{{k}}v_{{j}}+v_{{k}}x_
{{j}} \right) ^{2}}{-{a_{{j}}}^{2}+{a_{{k}}}^{2}}} \, , \,  k = 1 \dots N+1
$$
are conserved quantities (``integrals of motion").
Although this list includes $N+1$ conserved quantities, one checks that $\sum_{k=1}^{N+1} I_k = \sum_{k=1}^{N+1}v_k^2$; and this last sum is a conserved quantity for any thin film equation  (a Casimir for our setup).  
One checks that the Poisson bracket $\{I_j, I_k\} = 0$, that is, the functions are in involution.  It is also possible to obtain these constants of motion by considering the first integrals for geodesic flow on the ellipsoid (\cite{Tb}), and then taking the limit as the thickness of the ellipsoid goes to zero.
\end{subsection}
\end{section}
\begin{section}{``Nearly thin" double mirror systems and  TFE's:  a ``numerical analysis" perspective}
Suppose we have a surface $M$, a displacement function $f$ (which we assume to be strictly positive) and the nearby surface $M_\epsilon = \{ p + \epsilon f(p)N(p), p \in M_0\}$, where $\epsilon$ is small, fixed, and positive.
Suppose we have a point $p_0$ on $M$, and an initial unit pseudo-velocity vector $v_0$.  We can then do a ``double reflection" as in Figure 1, obtaining a new point and pseudo-velocity $p_1$ and $v_1$.    Alternatively, we can solve the differential equation system $(\dot v, \dot p) =   \chi(v,p)$ with initial conditions $v(0) = v_0, p(0) = p_0$ and ask for the values of $v(\epsilon)$ and $p(\epsilon)$, and ask how close these values are to $v_1$ and $p_1$.  One can think of the mapping $(v_0, p_0) \rightarrow (v_1, p_1)$ as a 
( not very efficient!) way of obtaining a discrete approximation to $v_1$ and $p_1$, just as one does in numerical analysis.  What sort of estimates can one make?

Such a result may be useful in a subsequent paper (see concluding remarks), but here we will just sketch the results associated to a very elementary example, which we believe exhibits the behavior of the more general problem.  The full details will have to appear in a sequel.

 We work in two dimensions, and $M$ is the straight line (one-dimensional flat template!) $p_2 = 0$.  Assume $p_0 = 0$,  $v(0)$ not horizontral and
$f(p) = a + b p + c p^2$.  Then one can do the following two calculations (greatly facilitated by symbolic calculation):  (i) Calculate, to third order in $\epsilon$, the reflected point $p_1$ and $v_1$;  and (ii) Calculate a series expansion in $\epsilon$, up to third order,  for the solution $v(\epsilon), p(\epsilon)$ of  the system $(\dot v, \dot p) =   \chi(v,p)$  with  $v(0) = v_0, p(0) = p_0$.  Comparing these two expansions, one sees that the difference is $O(\epsilon^3)$.  Typical arguments in numerical analysis (\cite{Dahl}) in numerical analysis show that, if the error associated with this single time-step is $O(\epsilon^3)$, then the difference between $(v_N, p_N)$ and $(v(N \epsilon), p(N \epsilon))$ is $O(\epsilon^2)$.  
\end{section}

\begin{section} {Bracket Structure}
Obviously, the set of all functions  of the form $H_f = \la v, N \ra f$ (as before, $f: M \rightarrow R$ ) forms a vector subspace of the set of all functions on $G_2$.  We ask how they behave with respect to the Poisson bracket associated with the symplectic form on $G_2$.  We record a couple of relevant formulas.  First, some notation.  As before, we have $H_f(v,p) = \la v, N(p) \ra f(p)$.    Let $\bf w$ be a vector field on $M$, then $H_{\bf w}(v,p) = \la v, \bf w \ra$. We have:
$$
\begin{array}{lcl}
\lb H_f, H_g \rb &=& H_{f \nabla g - g \nabla f}  \\
\lb H_{\bf w_1} , H_{\bf w_2} \rb  &=&  H_{[\bf w_1, \bf w_2]}  
\end{array}
$$
The proof of these formulas in the simple case where $M$ is the flat hyperplane $x_{n} = 0$, in standard Euclidean coordinates, is elementary.  We then observe that, in any coordinate representation, all of the quantities involved are first order, and invariant under translations and rotations; hence can be reduced to the simple case just discussed. 
Note that the second formula shows an isomorphism between the Lie algebra of vector fields along $M$, and their associated Hamiltonians.  Of course, if 
$P$ is a manifold and $T^* P$ is its cotangent bundle, then any vector field on $P$ lifts (locally) to a hamiltonian vector field on $T^*P$.  Since $G_2$ is (by definition) symplectomorphic  to
$T^* S^n$, it is not surprising that a similar statement holds in our situation; it is nice that the form of the equations works out so cleanly. 

It would be interesting to understand  the full Lie algebra generated by the Hamiltonians $H_f$.  

\end{section}

\begin{section}{Concluding Remarks and Related questions}
Of course, ray optics is derived from Maxwell's equations by taking the limit as frequency $\lambda$ becomes large.  In this paper we let the thickness $\epsilon$ of the double mirror system (thin film) go to zero {\it after} letting $\lambda$ approach infinity.  Clearly, it would be more realistic to have some coupling  between $\lambda$ and $\epsilon$ so that they go to infinity and zero respectively at the same time.  This is an important aspect of the problem which we wish to pursue in the future.

In \cite{Pei}, Pierone has a result which, translated into the language of our paper, says that for a positive displacement function $f$ and for $\epsilon$ small (but not zero), the billiard problem is transitive on configuration space:  given two points $p_0, p_1$ on $M$, there is a starting direction vector $v_0$ such that by going out along $v_0$ starting from $p_0$ , one lands at $p_1$ after sufficiently many reflections.  We are hoping that the considerations of this paper will allow for a simpler proof of Pierone's result,  by approximating the (discrete)  billiard trajectories by the (continuous) trajectories of $\chi_f$.  Hopefully, the considerations of Section 8 will prove useful here.

The variational problem associated with geodesics on a manifold is of course $\mathcal{L}=\int \sqrt{  \Sigma  \, g_{ij}\dot{ \gamma_i} \dot{ \gamma_j}} \, ds$.  As suggested by Section 7, one could instead consider the generalization $\mathcal{F}=\int \sqrt{ f^2 + \Sigma \, g_{ij}\dot{ \gamma_i } \dot{\gamma_j} } \, ds$, associated with a displacement function $f$.  What properties of geodesic flow are preserved, or can be preserved, by this perturbation of $\mathcal{L}$?  For example, if the geodesic equations are integrable, are there $f$ for which the Euler-Langrange equations  are integrable?  In a subsequent paper, we discuss such questions in the context of thin film equations associated to conics.

\end{section}
\begin{section}{Acknowledgements}
This paper would not exist without the catalyst of various communications with Sergei Tabachnikov, both in person and via email, to whom I owe great thanks.  The ICERM Conference on Integrability in June 2015 gave me the opportunity to discuss these ideas with the many visitors there, and I want to thank the organizers of the conference as well as ICERM itself for making it all pleasant and possible.  Various colleagues have discussed some of these ideas with me at length; in particular I want to thank Joel Langer of Case Western Reserve and Doug Wright of Drexel for their attention and advice.
\end{section} 

\bibliographystyle{unsrt} 
\bibliography{thinfilms}
\includegraphics{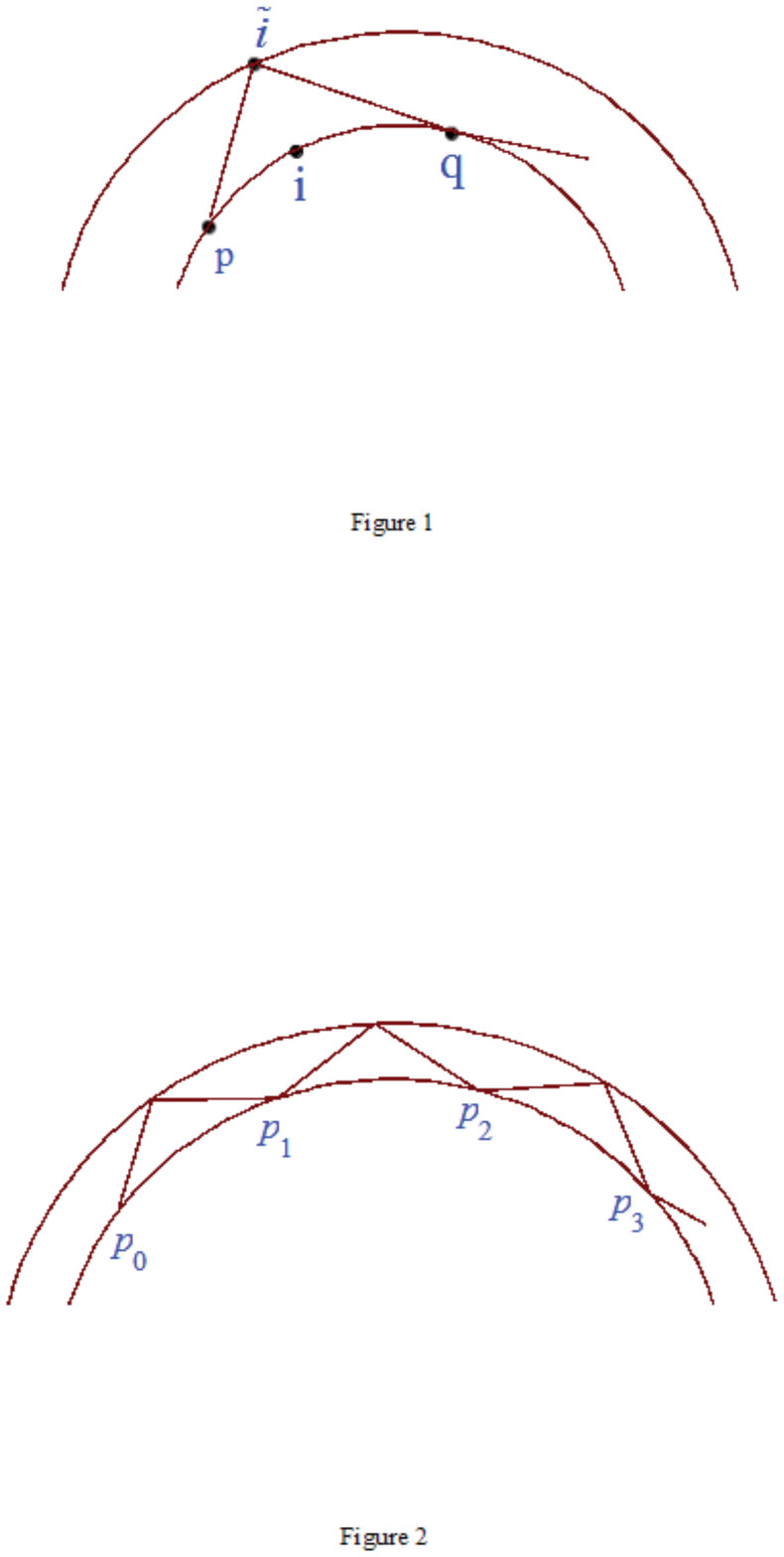}
\includegraphics{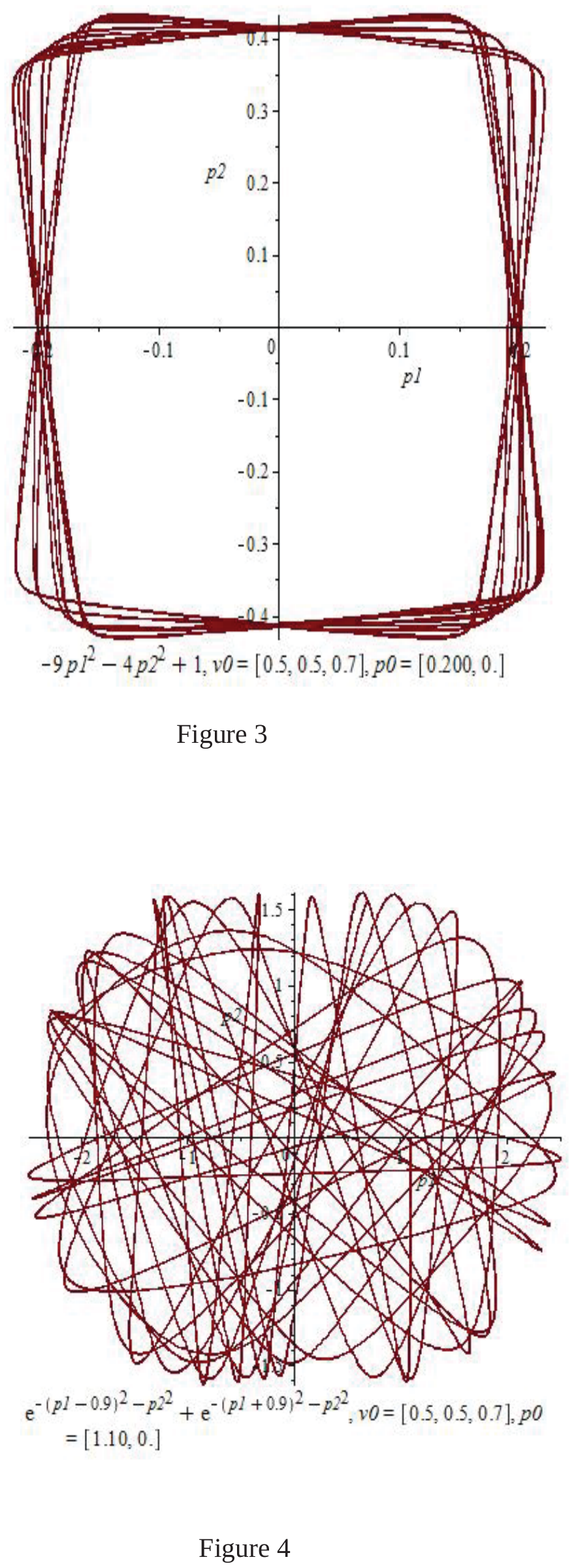}
\includegraphics{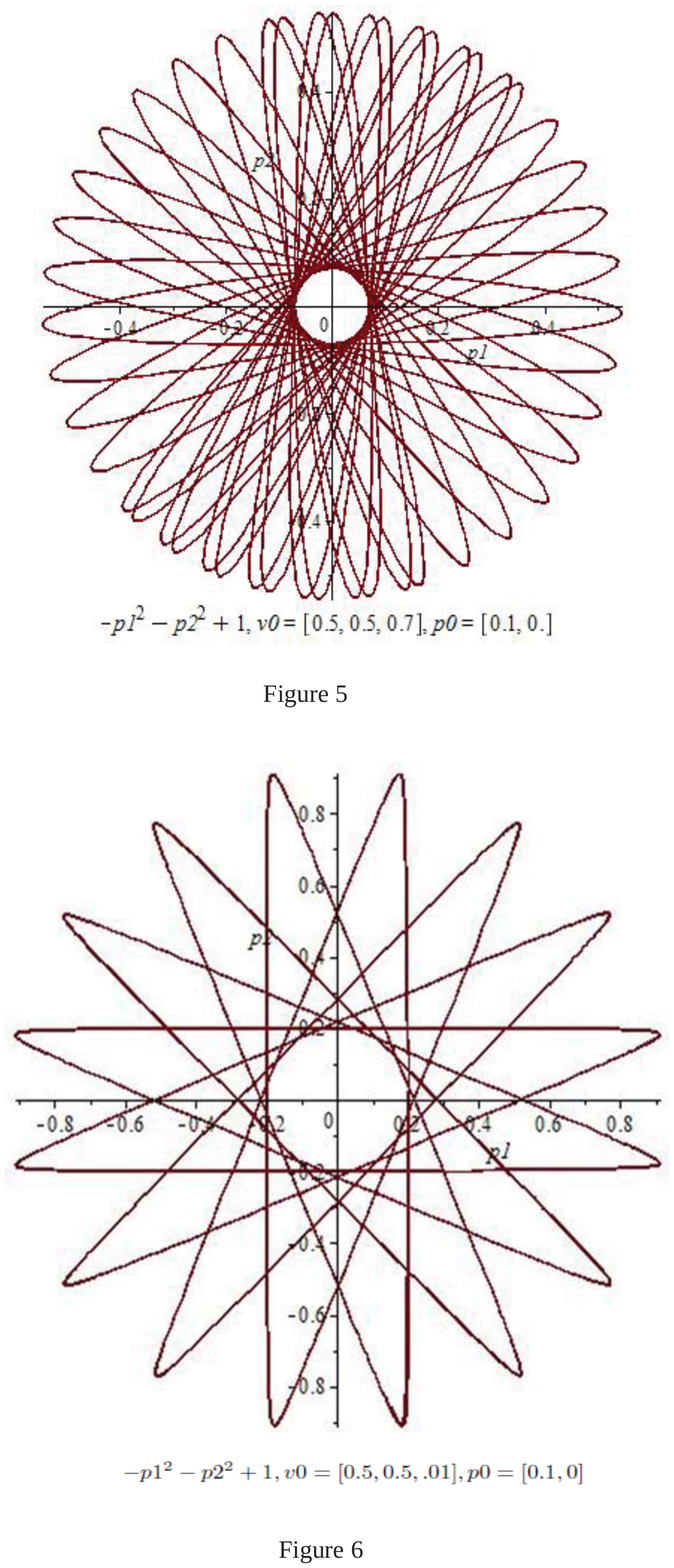}
\includegraphics{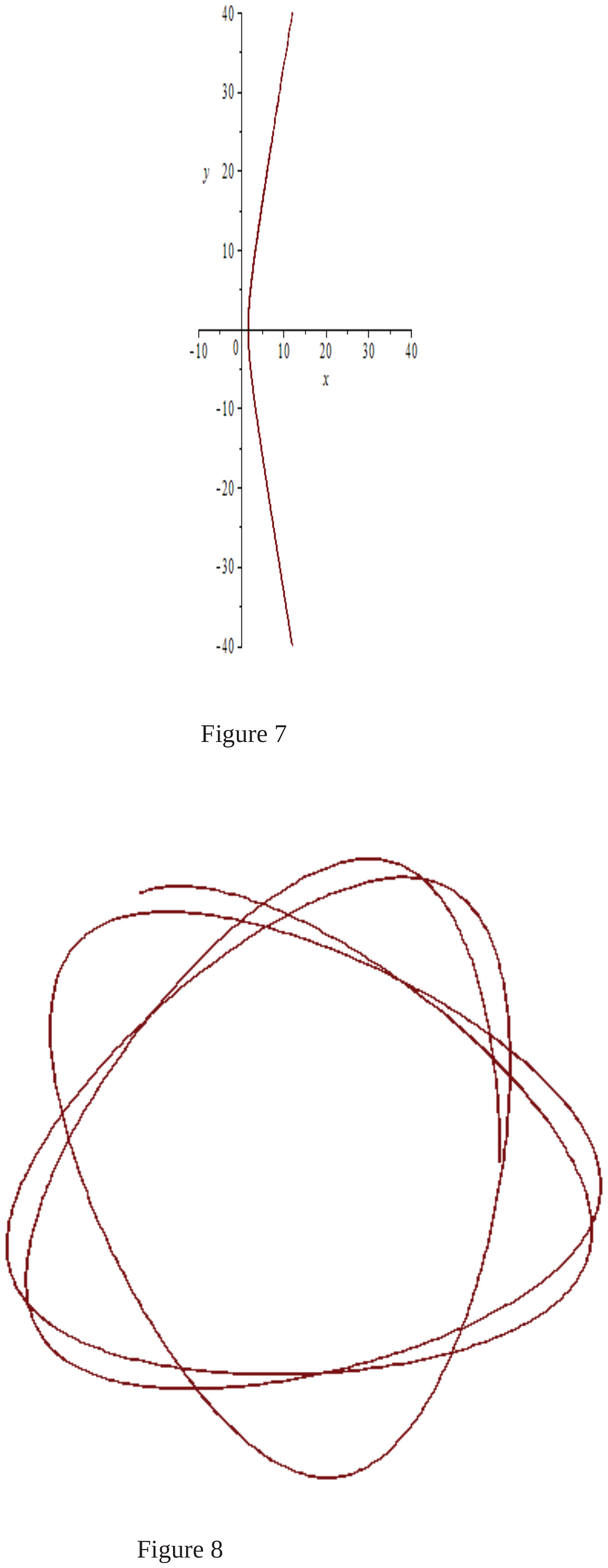}
\includegraphics{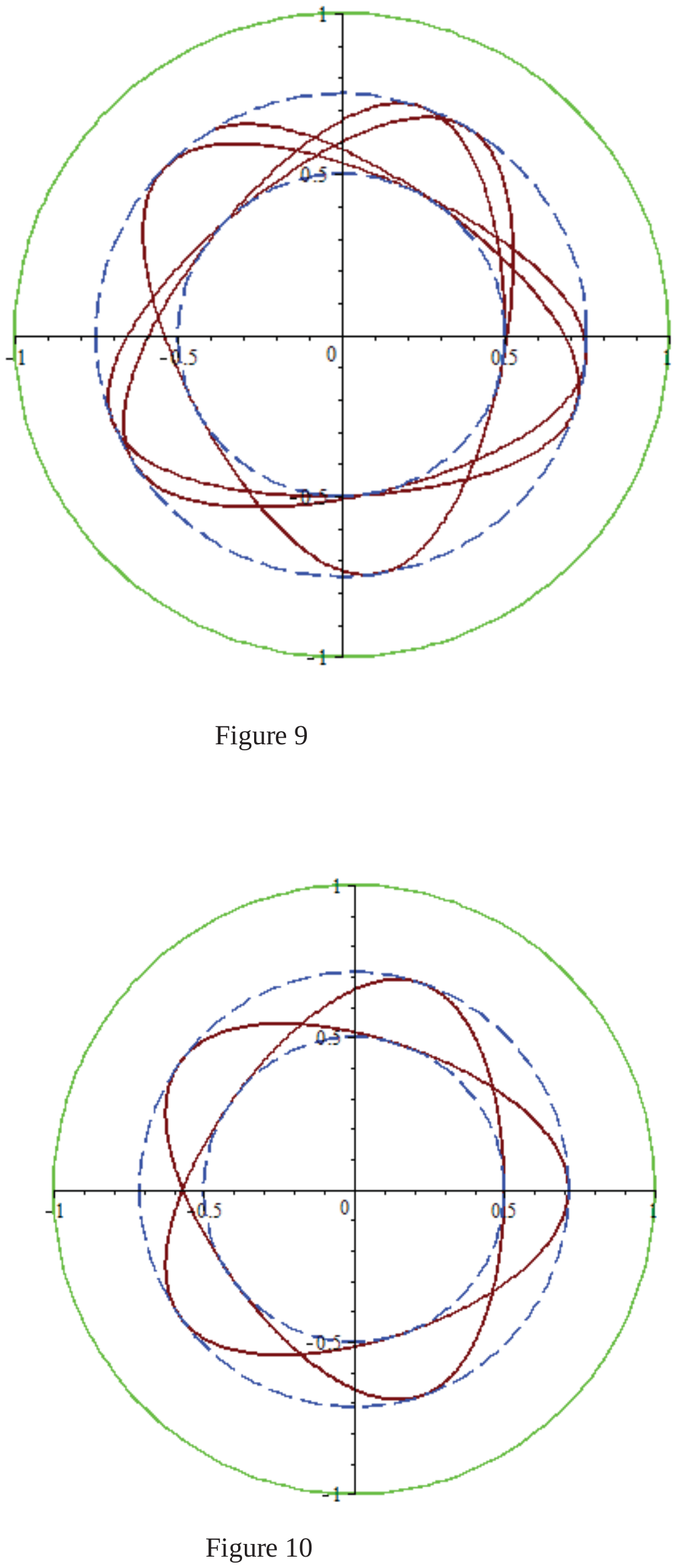}
\includegraphics{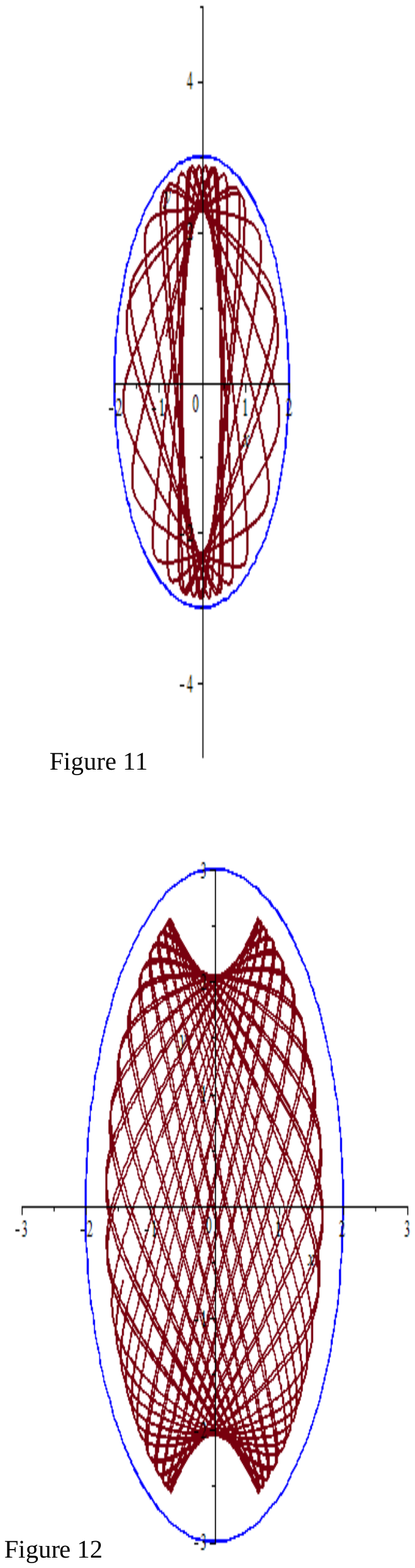}

\end{document}